\documentclass[apj]{emulateapj}

\begin{document}

\shorttitle{X-rays from Millisecond Pulsars in NGC 6397}
\shortauthors{Bogdanov et al.}

\title{A {\it Chandra X-ray Observatory} Study of PSR J1740--5340 and \\ Candidate Millisecond Pulsars in the Globular Cluster NGC 6397}

\author{Slavko Bogdanov\altaffilmark{1,2}, Maureen van den
  Berg\altaffilmark{3}, Craig O.~Heinke\altaffilmark{4}, Haldan
  N.~Cohn\altaffilmark{5}, \\ Phyllis M.~Lugger\altaffilmark{5}, Jonathan E.~Grindlay\altaffilmark{3}}

\altaffiltext{1}{Department of Physics, McGill University, 3600 University Street, Montreal, PQ H3A 2T8, Canada; bogdanov@physics.mcgill.ca}

\altaffiltext{2}{Canadian Institute for Advanced Research Junior Fellow}

\altaffiltext{3}{Harvard-Smithsonian Center for Astrophysics, 60 Garden Street,
Cambridge, MA 02138, U.S.A.}

\altaffiltext{4}{Department of Physics, University of Alberta, Edmonton, AB T6G 2G7, Canada}

\altaffiltext{5}{Department of Astronomy, Indiana University, 727 East Third Street, Bloomington, IN 470405, U.S.A.}

\begin{abstract}
  We present a deep \textit{Chandra X-ray Observatory} study of the
  peculiar binary radio millisecond pulsar PSR J1740--5340 and
  candidate millisecond pulsars (MSPs) in the globular cluster NGC
  6397.  The X-rays from PSR J1740--5340 appear to be non-thermal and
  exhibit variability at the binary period.  These properties suggest
  the presence of a relativistic intrabinary shock formed due to
  interaction of a relativistic rotation-powered pulsar wind and
  outflow from the unusual ``red-straggler/sub-subgiant'' companion.
  We find the X-ray source U18 to show similar X-ray and optical
  properties to those of PSR J1740--5340, making it a strong MSP
  candidate.  It exhibits variability on timescales from hours to
  years, also consistent with an intrabinary shock origin of its X-ray
  emission. The unprecedented depth of the X-ray data allows us to
  conduct a complete census of MSPs in NGC 6397. Based on the
  properties of the present sample of X-ray--detected MSPs in the
  Galaxy we find that NGC 6397 probably hosts no more than 6
  MSPs.

\end{abstract}


\section{INTRODUCTION}

Globular clusters are well known for their abundance of
rotation-powered millisecond pulsars \citep{Camilo00, Freire03,
  Ran05}\footnote{For an up-to-date list of all known globular cluster
  MSPs see http://www.naic.edu/~pfreire/GCpsr.html.}.  These objects
are believed to have been produced by the evolution of low-mass X-ray
binaries (LMXBs) that were themselves produced through dynamical
interactions of binaries and neutron stars.  The superb sub-arcsecond
X-ray imaging capability of the \textit{Chandra X-Ray Observatory} has
allowed X-ray studies of MSPs in the dense cores of globular clusters
for the first time.  In addition to establishing the X-ray properties
of the Galactic population of MSPs, X-ray studies of cluster MSPs
offer constraints on stellar and binary evolution and the internal
dynamical evolution of globular clusters.  To date, X-ray
counterparts of these systems have been detected with
\textit{Chandra} in 47 Tuc \citep{Grind02,Bog06}, NGC 6397
\citep[][and this work]{Grind02}, M28 \citep[][S.~Bogdanov et al. in
preparation]{Beck03}, M4 \citep{Bass04a}, NGC 6752 \citep{DAmico02},
M71 \citep{Els08}, and Ter 5 \citep{Hein06}.

NGC 6397 is the nearest apparently core-collapsed and the second
closest globular cluster, with $D\approx2.4$ kpc \citep[][and
references therein]{Han07,Strick09}.  It hosts one known MSP, PSR
J1740--5340 \citep{DAmico01}, which is bound to an unusual
``red-straggler'' or ``sub-subgiant''\footnote{The terms red straggler
  and sub-subgiant are used for cluster stars that lie in a location
  below the base of the red giant branch that is difficult to
  understand in the context of standard evolution scenarios for single
  stars or binaries in a dynamically inactive environment
  \citep{Oro03}.} companion in a 32.5-hour binary orbit
\citep{Ferr03}. This is at odds with the standard MSP formation theory
that predicts either a white dwarf or a very-low-mass ($\sim$0.03
M$_{\odot}$) degenerate companion \citep[see][for a review]{Bhatt91},
implying that either this binary has just emerged from the
``recycling'' phase in a LMXB \citep{Bur02} or has been involved in a
close dynamical binary-binary encounter in which the original
companion to the pulsar was exchanged for the current one \citep[][and
references therein]{Cam05}.

Observations of MSPs, both in the field of the Galaxy and in globular
clusters, have revealed that their X-ray emission can be of thermal
and/or non-thermal character \citep[see][for a
review]{Zavlin07,Bog08a}.  In many MSPs, the X-rays appear to be
predominantly due to surface emission from the magnetic polar caps of
the neutron star \citep{Bog06,Zavlin06,Bog09}. Non-thermal pulsed
X-ray emission, on the other hand, can arise from particle
acceleration processes in the pulsar magnetosphere as seen from the
most energetic MSPs \citep[see, e.g.][]{Rut04}. Alternatively,
non-thermal X-rays can be produced via interaction of the
rotation-powered pulsar wind with the ambient interstellar medium, as
seen for PSR B1957+20 \citep{Stap03} and J2124--3358 \citep{Hui06}, or
material from a close binary companion as in PSR J0024--7204W in 47
Tuc \citep{Bog05}.

%
%
%
\begin{figure*}[!t]
\begin{center}
\includegraphics[width=0.85\textwidth]{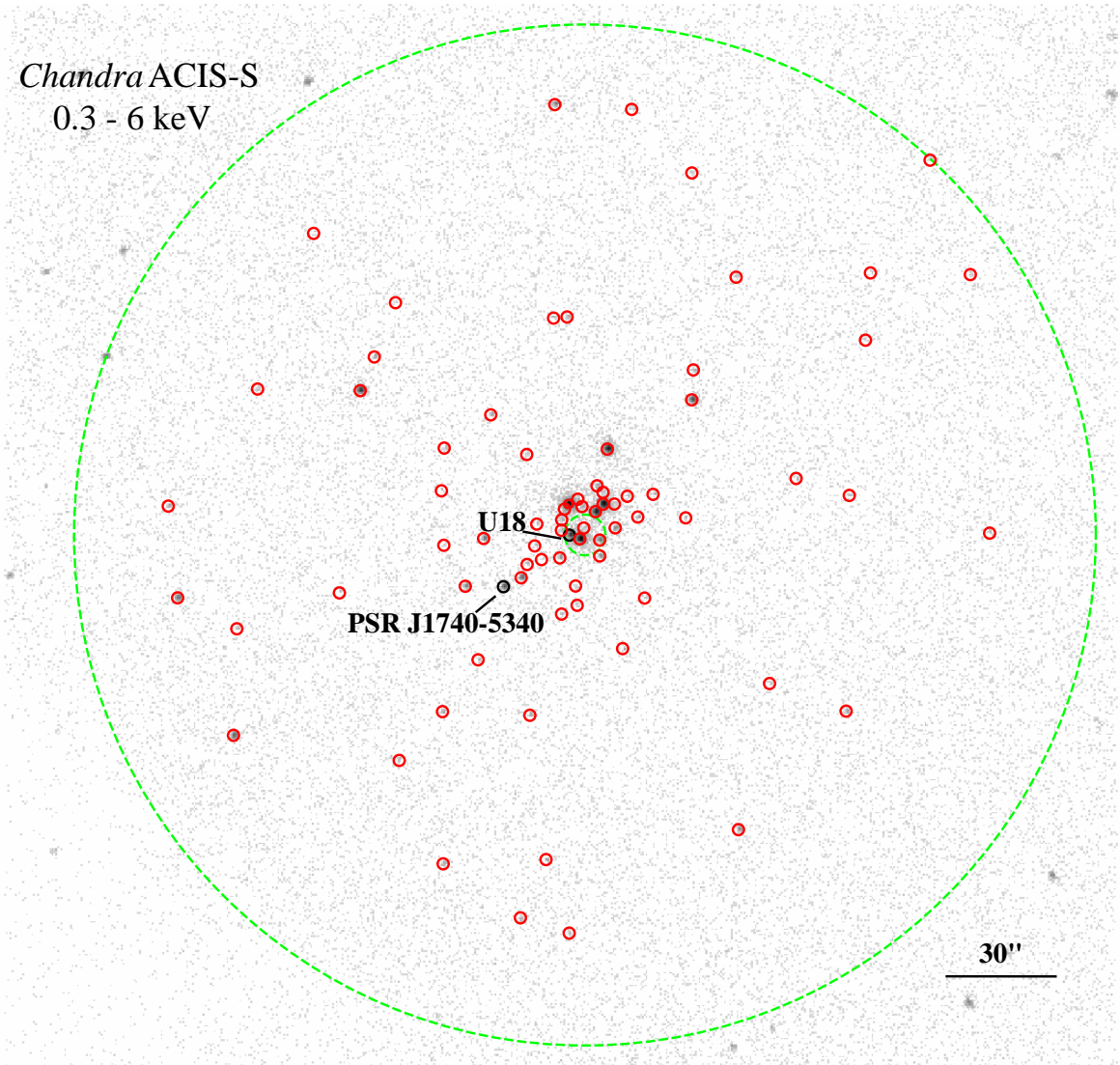}
\end{center}
\caption{Coadded \textit{Chandra X-ray Observatory} 295.1-ks ACIS-S3
  image of the core of the globular cluster NGC 6397 in the 0.3--6 keV
  band. The $1.5\arcsec$ circles are centered on the 79 X-ray sources
  detected within the $2.3\arcmin$ half-mass radius of the cluster
  (outer dashed circle), with the positions of PSR J1740--5340 (U12)
  and CXOGlb J174041.6--534027 (U18) marked. The inner dashed circle
  shows the $5.5\arcsec$ core radius of the cluster. The grayscale
  corresponds to number of counts increasing logarithmically from 0
  (\textit{white}) to 3515 (\textit{black}). North is up and east is
  to the left.}
\end{figure*}

In this paper, we present \textit{Chandra} deep imaging spectroscopic
observations of NGC 6397, with a particular focus on PSR J1740--5340
and CXOGlb J174041.6--534027 (U18), a plausible MSP candidate. We also
investigate the X-ray source population of NGC 6397 in an attempt to
identify plausible MSP candidates and constrain the MSP content of
this cluster.  The work is organized as follows. In \S 2 we describe
the data reduction and analysis procedures. In \S 3 we focus on the
properties of PSR J1740--5340, while in \S 4 we investigate the MSP
candidate U18. In \S5 we attempt to place interesting constraints on
the total number of MSPs in the cluster based on the available X-ray
and optical data. We present a discussion in \S6 and conclusions in
\S7.

\section{DATA REDUCTION AND ANALYSIS}

The \textit{Chandra} data discussed herein were acquired during two
separate observations in Cycle 8, on 2007 June 22 (ObsID 7461) and
2007 July 16 (ObsID 7460) for 90 and 160 ks, respectively.  In both
cases the ACIS S3 chip in VFAINT telemetry mode was at the focus. We
also make use of two Cycle 3 ACIS-S observations of 28.5 ks and 27 ks
(ObsIDs 2668 and 2669, respectively), and a single Cycle 1 ACIS-I
(ObsID 79) observation of 49 ks \citep{Grind01}, all acquired in FAINT
mode. Table 1 summarizes all the observations used in this work.

The data re-processing, reduction, and analysis were performed using
CIAO\footnote{Chandra Interactive Analysis of Observations, available
  at http://cxc.harvard.edu/ciao/} 4.0.  Starting from the level 1
data products, we first removed pixel randomization from the standard
pipeline processing in order to aid in source disentanglement in the
dense cluster core.  In addition, for the purposes of faint source
detection we applied the background cleaning algorithm for the
data taken in VFAINT observing mode. However, as this procedure tends
to discard real source counts for relatively bright sources, the
background-cleaned data were not used for the spectral analyses
discussed in \S3 and \S4 but were used for the population study in \S5.
Before coadding the four ACIS-S images, the individual aspect
solutions were reprojected relative to the longest observation (ObsID
7460) using the brightest sources in the cluster in order to correct
for any differences in the absolute astrometry between the four
exposures.

%
%

\begin{figure}[t]
\begin{center}
\includegraphics[angle=270,width=0.45\textwidth]{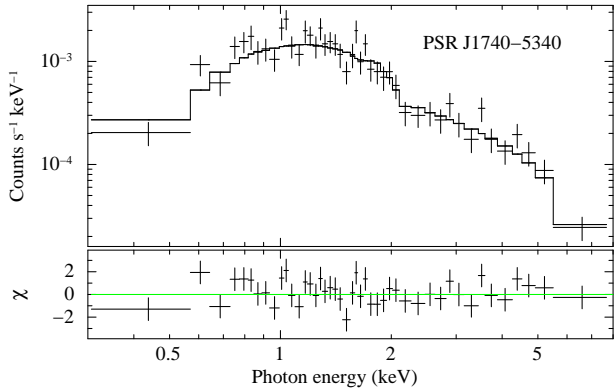}
\caption{Phase-integrated stacked spectrum of PSR J1740--5340 from all
  four \textit{Chandra} ACIS-S observations, fitted with an absorbed
  powerlaw spectrum. The lower panel shows the best fit residuals. See
  the text and Table 2 for the best fit parameters.}
\end{center}
\end{figure}

X-ray source detection was performed with the {\tt wavdetect} tool in
CIAO. In the crowded cluster core, we also used the
PWDetect\footnote{PWDetect has been developed by scientists at
  Osservatorio Astronomico di Palermo G.~S.~Vaiana thanks to Italian
  CNAA and MURST (COFIN) grants.} script \citep{Dam97a,Dam97b}, which
tends to be more effective at identifying faint sources near much
brighter sources. The resulting source positions were refined using
the IDL tool {\tt acis\_extract}. The 95\% confidence positional
uncertainty radius for each source was computed using the empirical
relation given by Equation (5) in \citet{Hong05}.  In total, 79
sources are detected within the 2.33' half-mass radius of NGC 6397
(see Table 3). Compared to the source list from \citet{Grind01}, based
on the single ACIS-I observation, and \citet{Grind05}, based on the
ACIS-I and two short ACIS-S observations, we find that all sources
except U20, U64, U71, U72, U74, U78, and U85 are detected in the
combined deep ACIS-S image.

For the purposes of optical matching of the X-ray sources (see \S5)
using the set of \textit{Hubble Space Telescope} Advanced Camera for
Surveys Wide Field Channel observations (GO-10257) in F435W (B), F625W
(R), and F658N (H$\alpha$), we computed a boresight correction for the
\textit{Chandra} source coordinates (see H.~Cohn, in preparation, for
details).  The resulting shift of the \textit{Chandra} coordinates is
$\Delta \alpha = -0.19\arcsec \pm 0.02\arcsec$ $\Delta \delta =
0.18\arcsec \pm 0.02\arcsec$. The X-ray position based on this
correction for PSR J1740--5340 agrees with the optical position
\citep{Bass04b} to $0.01\arcsec$ in $\alpha$ and $0.03\arcsec$ in
$\delta$.  While the \textit{Chandra} ACIS-S mosaic encompasses the
entire half-mass radius of the cluster, the \textit{HST} ACS/WFC
mosaic provides complete coverage out to a radius of $\sim$1$\farcs$5
from the cluster center, partial coverage in two or more exposures out
to $\sim$$2\farcs5$, and partial coverage in at least one exposure out
to $2\farcs9$. As a result, the sources U5, U16, U77, and U84, lie
outside the $R$ field making their classification difficult due to
the absence of $B−R$ and H$\alpha-R$ color measurements. However,
\citet{Kal06} have shown that the optical counterparts of U5 and U77,
V30 and V36 respectively, are variable using ground-based photometry
and thus are probably active binaries \citep{Kal06}.

Net counts and spectra for each X-ray source were extracted using {\tt
  acis\_extract} from polygonal regions sized to contain 90\% of the
total energy at 1.5 keV.  For spectroscopy, the extracted source
counts in the 0.3--8 keV range were then grouped in energy bins so as
to ensure at least 15 counts per bin. The X-ray spectral analyses of
PSR J1740--5340 and U18 were carried out using the
XSPEC\footnote{Available at
  http://heasarc.nasa.gov/docs/xanadu/xspec/index.html} package.  For
the variability analysis, the photon arrival times were first reduced
to the solar system barycenter, using the CIAO tool {\tt axbary}.

\begin{deluxetable}{lccc}
\tabletypesize{\small} 
\tablecolumns{5} 
\tablewidth{0pc}
\tablecaption{Chandra Observations of NGC 6397}
\tablehead{ \colhead{Telescope/} &\colhead{Epoch of} & \colhead{Observation} & \colhead{Exposure} \\
\colhead{Instrument} & \colhead{Observation} & \colhead{ID} &
\colhead{Time (ks)}}
\startdata
Chandra/ACIS-I & 2000 Jul 31 & 79 & 49.0 \\
Chandra/ACIS-S & 2002 May 13 & 2668 & 28.5 \\
Chandra/ACIS-S & 2002 May 15 & 2669 & 27.0 \\
Chandra/ACIS-S & 2007 Jun 22 & 7461 & 90.0 \\
Chandra/ACIS-S & 2007 Jul 16 & 7460 & 160.0
\enddata
\end{deluxetable}

\section{PSR J1740--5340}

\citet{Grind01,Grind02} have found the X-ray counterpart of the PSR
J1740--5340 system (see Figure 1) to be a moderately luminous
($L_X\sim10^{31}$ ergs s$^{-1}$) and relatively hard X-ray source
($\Gamma\sim1.4$), compared to most MSPs. Indeed, the combined
\textit{Chandra} ACIS-S spectrum (Figure 2) is well described by a pure
non-thermal model with power-law photon index $\Gamma=1.73\pm0.08$,
hydrogen column density of $N_H=(2.19^{+0.22}_{-0.25})\times10^{21}$
cm$^{-2}$, and unabsorbed X-ray flux $F_X=(3.2\pm0.4)\times10^{-14}$
ergs cm$^2$ s$^{-1}$ (0.3--8 keV) with $\chi_{\nu}^2=1.14$ for 42
degrees of freedom. All uncertainties quoted are 1$\sigma$. For an
assumed distance of 2.4 kpc to NGC 6397, the implied X-ray luminosity
in the 0.3--8 keV band is $L_X=2.2\times10^{31}$ ergs s$^{-1}$.

We also applied a two-temperature thermal (blackbody or H atmosphere)
model, which provides a good description of the X-ray spectra of
several nearby MSPs \citep{Zavlin06, Bog09}. However, this model fits
the phase-averaged spectrum of PSR J1740--5340 poorly.  The spectrum
of PSR J1740--5340 likely contains a soft thermal component (with
$T_{\rm eff}\sim10^6$ K and $L_{X}$ of order a few$\times10^{30}$ ergs
s$^{-1}$) originating from the hot magnetic polar caps of the MSP, as
seen in many X-ray--detected MSPs \citep[e.g. PSR J0437--4715 and most
MSPs in 47 Tuc, see][]{Zavlin06,Bog06}. To investigate this
possibility, we first introduce a single blackbody component to the
spectral model. Fitting a composite powerlaw plus blackbody model
yields $\Gamma=1.56^{+0.18}_{-0.23}$, $N_{\rm
  H}=(2.6^{+1.8}_{-0.8})\times10^{21}$ cm$^{-2}$,
$kT=0.19^{+0.09}_{0.04}$ K, and $R_{\rm eff}=0.15^{+0.09}_{-0.13}$ km
with $\chi_{\nu}^2=1.1$ for 40 degrees of freedom.  Although an F-test
indicates that the addition of a thermal component is not
statistically warranted by the data, the best fit values for $T_{\rm
  eff}$ and $R_{\rm eff}$ are similar to those obtained for the 47 Tuc
MSPs so this component may in fact be genuine. A composite model
consisting of a powerlaw plus two thermal components yields good fits
to the spectrum, although the limited photon statistics do not allow
meanigful constraints for a three-component model.  In principle, the
presence of surface thermal emission could be determined by way of
high time-resolution X-ray observations, which would reveal any
thermal pulsations. Unfortunately, the 3.2-second time resolution of
the ACIS observations precludes such an investigation with the present
dataset.

The X-ray emission from the PSR J1740--5340 system could also
originate from a thermal plasma within the binary, possibly from the
active corona of the secondary star or the material responsible for
the radio eclipses. To test this possibility, we used the {\tt vmekal}
thermal plasma model in XSPEC, with metal abundances set to values
representative of the stars in NGC 6397 \citep{Cast00,Grat01}. This model
also reproduces the observed spectral shape rather well, with best fit
parameters $N_{\rm H}=(1.80^{+0.25}_{-0.21})\times10^{21}$ cm$^{-2}$,
$kT=6.2^{+1.4}_{-1.1}$ keV, and $F_X=(2.70^{+0.29}_{-0.36})\times
10^{-14}$ ergs cm$^2$ s$^{-1}$ (0.3--8 keV), with $\chi_{\nu}^2=1.2$
for 47 degrees of freedom. As discussed in \S6.1, although a thermal
plasma model is consistent with the observed X-ray emission from the
PSR J1740--5340 binary, a predominantly non-thermal origin of the
X-rays is more likely.

%
%
\begin{figure}[t]
\begin{center}
\includegraphics[width=0.41\textwidth]{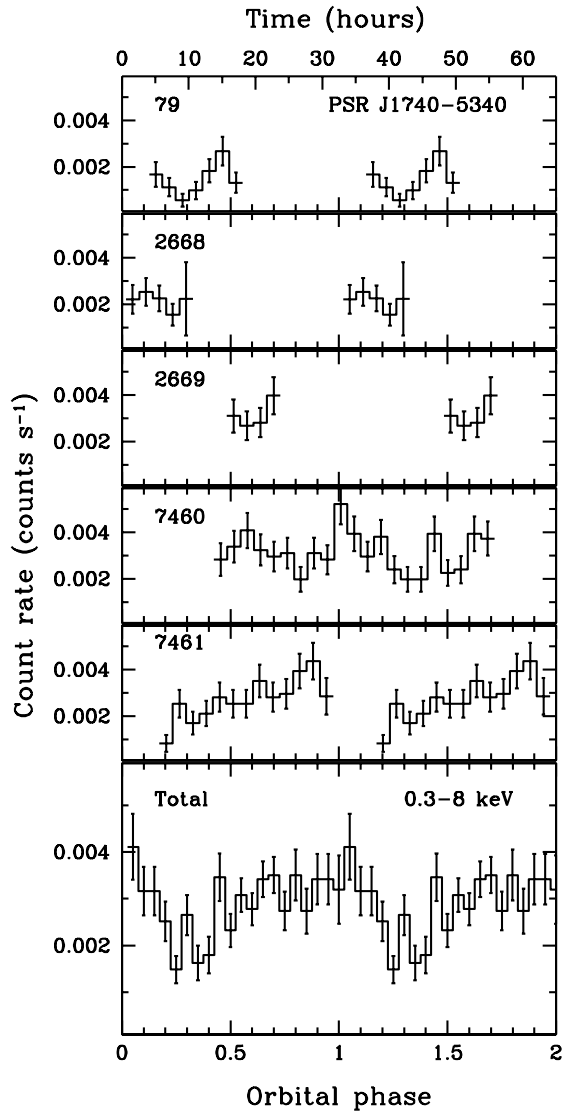}
\end{center}
\caption{Lightcurves of PSR J1740--5340 folded at the pulsar's binary
  period for each \textit{Chandra} ACIS observation. The bottom panel
  shows the combined folded lightcurve from the four ACIS-S
  observations, with the count rate corrected for the
  non-uniform exposure across the orbit. The phase is defined based on
  the radio pulsar timing convention in which superior conjunction
  occurs at $\phi=0.25$. Two orbital cycles are shown for clarity.}
\end{figure}

Based on the ACIS-I observation of NGC 6397, \citet{Grind02}
reported evidence for a gradual increase (by a factor of 2 in total)
in the X-ray count rate of PSR J1740--5340. However, due to the
limited phase coverage and photon statistics, no conclusive statements
could be made regarding variability.  Figure 3 shows the X-ray count
rate from the pulsar as a function of binary phase, based on the radio
timing ephemeris from \citet{DAmico01}, of all \textit{Chandra}
observations of PSR J1740--5340.  Although the individual lightcurves
are suggestive of flux modulations, with generally lower count rates
around the radio eclipse phases (0.05--0.45), a Kolmogorov-Smirnov
(K-S) test on the unbinned lightcurves within each observation reveals
no statistically significant variability. In addition, the same test
indicates variability only at a 1.7$\sigma$ confidence level over the
entire set of ACIS-S observations. On the other hand, based on Poisson
statistics, we find that the count rate minimum that occurs near
$\phi=0.25$ (see bottom panel of Fig.~3) for the lightcurve folded at
the binary period and grouped in 20--40 bins, deviates by $\sim$3.8$\sigma$
from what is expected from a constant source.  A $\chi^2$ test on the
same lightcurve indicates a 98.7\% probability of variability. There
is also marginal evidence (at $\sim$95\% confidence) for spectral
variability, with an apparent softening of the X-ray emission around
$\phi=0.25$.  Due to the long binary period, the ACIS-S data cover
less than 3 full binary orbits so it is not clear whether this
variability is truly periodic and stable over many orbits.
Interestingly, the folded lightcurve exhibits the same characteristic
shape as that of PSR J0024--7204W in 47 Tuc \citep{Bog05}, with a
minimum around superior conjunction, roughly coincident with the radio
eclipses. The spectral and temporal similarities of the two systems
point to an intrabinary shock origin of the X-rays from PSR
J1740--5340.

%
%
\begin{figure}[t]
\begin{center}
\includegraphics[angle=270,width=0.45\textwidth]{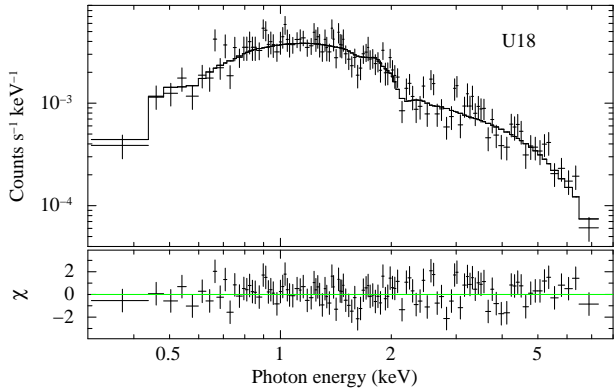}
\caption{Total \textit{Chandra} ACIS-S spectrum of the X-ray source
  CXOGlb J174041.6--534027 (U18) fitted with a pure powerlaw
  spectrum. The bottom panel shows the best fit residuals. See text
  and Table 2 for best fit parameters.}
\end{center}
\end{figure}

\section{U18: A MILLISECOND PULSAR CANDIDATE}

The X-ray source CXOGlb J174041.6--534027 (hereafter U18) is
positionally coincident with a peculiar ``red-straggler''
star\footnote{We note that the proper motions of the optical
  counterparts to U12 and U18 are consistent with cluster membership
  (see H.~Cohn, et al., in preparation). Therefore, their anomalous
  locations in the cluster optical color-magnitude diagram cannot be
  explained by them being foreground or background objects.}
\citep{Grind01}. In addition, it is known to have a relatively hard
spectrum making it a plausible candidate for a MSP given the
similarities in X-ray and optical properties with PSR J1740--5340.  A
fit of a power-law model to the combined ACIS-S spectrum of U18 gives
$\Gamma=1.38^{+0.05}_{-0.04}$,
$N_H=(1.4^{+0.12}_{-0.11})\times10^{21}$ cm$^{-2}$ with
$\chi_{\nu}^2=1.0$ for 114 degrees of freedom. The derived unabsorbed
flux of $F_X=9.7\times10^{-14}$ ergs cm$^{-2}$ s$^{-1}$ (0.3--8 keV)
implies a luminosity of $L_X=6.7\times10^{31}$ ergs s$^{-1}$ for
$D=2.4$ kpc. Figure 4 shows the spectrum of U18 in the 0.3--8 keV band
and the best fit PL model.

As with PSR J1740--5340, we also conducted fits using a two component
(thermal plus non-thermal) model. The resulting best fit values are
$\Gamma=1.29^{+0.05}_{-0.12}$, $kT=0.19^{+0.03}_{-0.03}$ keV, $R_{\rm
  eff}=0.18^{+0.13}_{-0.14}$,
$N_H=(1.70^{+0.05}_{-0.07})\times10^{21}$ cm$^{-2}$, with
$\chi_{\nu}^2=0.98$ for 112 degrees of freedom. The unabsorbed flux in
the 0.3--8 keV band is $1.07\times10^{-13}$ ergs cm$^{-2}$ s$^{-1}$,
which corresponds to an X-ray luminosity of $L_X=7.4\times10^{31}$
ergs s$^{-1}$ for $D=2.4$ kpc. By analogy with the nearest known MSPs
\citep{Zavlin06,Bog09}, we also apply a model with a powerlaw plus two
thermal components. This model is consistent with the observed
spectrum as well, although the relatively high column density along
the line of sight, which absorbs a large fraction of source photons
below $\sim$0.5 keV, makes it difficult to reliably constrain the
softer thermal component.

A thermal plasma model ({\tt vmekal} with the same assumed abundances
as for PSR J1740--5340) also provides a good description of the
spectrum of U18. The best fit parameters in this case are
$N_H=(1.53^{+0.14}_{-0.13})\times10^{21}$ cm$^{-2}$,
$kT=16.8^{+4.7}_{-3.4}$ keV, and $F_X=9.0\times10^{-14}$ ergs
cm$^{-2}$ s$^{-1}$ (0.3--8 keV), with $\chi_{\nu}^2=1.06$ for 128
degrees of freedom.

\begin{deluxetable}{lcc}
\tabletypesize{\footnotesize} 
\tablecolumns{3} 
\tablewidth{0pc}
\tablecaption{Spectral fits for PSR J1740--5340 and U18}
\tablehead{ \colhead{} & \colhead{PSR J1740--5340} & \colhead{CXOGlb J174041.7--534027} \\ \colhead{Parameter} & \colhead{(U12)} & \colhead{(U18)} }
\startdata
Powerlaw & & \\
\hline
$N_{\rm H}$ ($10^{21}$ cm$^{-2}$) & $2.19^{+0.22}_{-0.25}$ & $1.40^{+0.12}_{-0.11}$ \\
$\Gamma$ & $1.73^{+0.08}_{-0.08}$ & $1.38^{+0.05}_{-0.04}$ \\
$F_X$\tablenotemark{a} (0.3--8 keV) & $3.22^{+0.37}_{-0.36}$ & $9.72^{+0.07}_{-0.02}$ \\
$\chi^2_{\nu}/{\rm dof}$ & 1.14/42 & 1.00/114 \\
\hline
Powerlaw+Blackbody & & \\
\hline
$N_{\rm H}$ ($10^{21}$ cm$^{-2}$)  & $2.58^{+1.80}_{-0.80}$ & $1.70^{+0.05}_{-0.07}$ \\
$\Gamma$ & $1.56^{+0.18}_{-0.23}$ & $1.29^{+0.05}_{-0.12}$ \\
$kT$ (keV) & $0.19^{+0.09}_{-0.04}$ & $0.19^{+0.03}_{-0.03}$ \\
$R_{\rm eff}$\tablenotemark{b} (km) & $0.15^{+0.09}_{-0.13}$ &  $0.18^{+0.13}_{-0.14}$ \\
$F_X$\tablenotemark{a} (0.3--8 keV) & $3.47^{+0.39}_{-1.66}$ & $10.71^{+0.08}_{-0.03}$\\
$\chi^2_{\nu}/{\rm dof}$ & $1.11/40$ & $0.98/112$ \\
\hline
VMEKAL & & \\
\hline
$N_{\rm H}$($10^{21}$ cm$^{-2}$) & $1.80^{+0.25}_{-0.21}$ & $1.53^{+0.14}_{-0.13}$  \\
$kT$ (keV) & $6.2^{+1.4}_{-1.1}$ & $16.8^{+4.7}_{-3.4}$ \\
$F_X$\tablenotemark{a} (0.3--8 keV) & $2.70^{+0.29}_{-0.36}$ & $9.04^{+0.57}_{-0.71}$  \\
$\chi^2_{\nu}/{\rm dof}$ & $1.19/47$ &  $1.06/128$
\enddata
\tablenotetext{a}{Unabsorbed flux in units of $10^{-14}$ ergs cm$^{-2}$ s$^{-1}$ in the 0.3--8 keV band.}
\tablenotetext{b}{Effective blackbody emission radius assuming a distance of 2.4 kpc to NGC 6397.}
\end{deluxetable}

It is obvious from Figure 5 that the X-ray flux from U18 exhibits
variability by as much as a factor of $\sim$4 on timescales spanning
from hours to years.  Indeed, a K-S test on the concatenated set of
ACIS-S observations indicates a $3.5\times10^{-15}$ probability of the
observed photons coming from a steady flux distribution. Moreover, the
photon arrival times from U18 within observations 2668, 2669, 7460,
and 7461 deviate by 3$\sigma$, 2.3$\sigma$ , 3$\sigma$, and 4$\sigma$
from a constant distribution, based on a K-S test.

\citet{Kal06} have reported the likely detection of periodicity,
possibly due to ellipsoidal variations, with $P=1.3$ days in the
source V31, which they associate with U18.  It should be noted,
however, that the position of V31 lies at $\sim$2.1$\sigma$ from U18
after boresight correction. Moreover, V31 does not correspond to the
red-straggler optical counterpart of U18 reported by \citet{Grind01},
which is only $\sim$0.4$\sigma$ away. We have folded the set of four
\textit{Chandra} ACIS-S observations at the period of V31 but find no
statistically significant variability. This implies that V31 is most
likely not the actual optical counterpart of U18.

%
%
\begin{figure}[t]
\begin{center}
\includegraphics[width=0.42\textwidth]{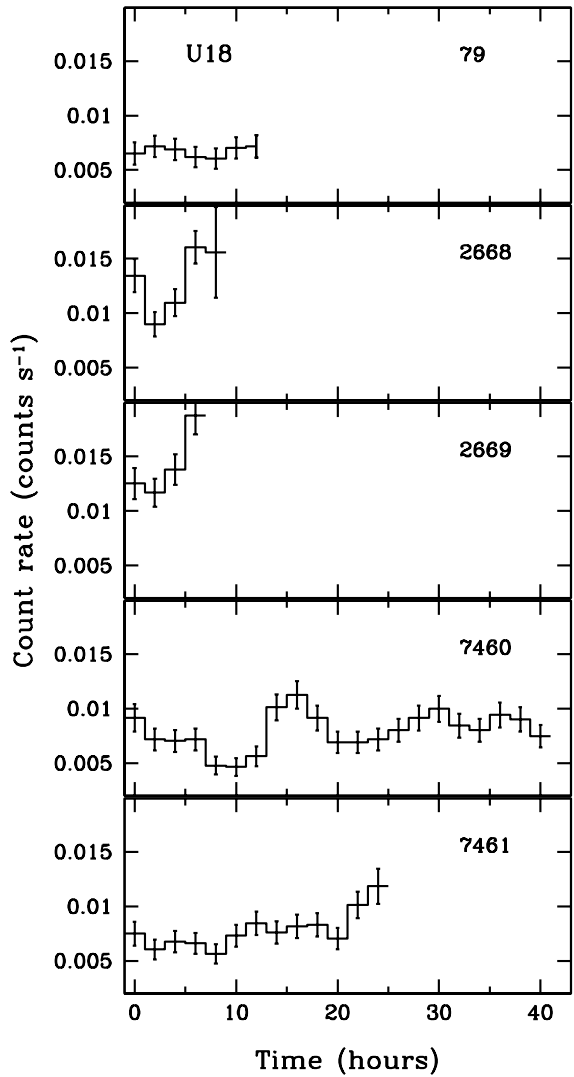}
\caption{X-ray lightcurves of CXOGlb J (U18) in the 0.3--8 keV band from all
  \textit{Chandra} ACIS observations binned in 1-hour
  intervals. Time zero corresponds to the start of each observation.}
\end{center}
\end{figure}

\section{AN X-RAY CENSUS OF MILLISECOND PULSARS  IN NGC 6397}

As noted previously, several nearby MSPs in the field of the Galaxy
\citep{Zavlin06,Bog09} and most MSPs in the globular cluster 47 Tuc
\citep{Bog06} appear to be predominantly soft, thermal X-ray sources
due to their hot polar caps. As demonstrated by \citet{Bog08} (see in
particular their Figure 1), the (nearly) antipodal geometry of the
polar caps and the effects of light bending guarantee that this
surface emission is observed for any combination of magnetic
inclinations and viewing angles. In addition, the difference in
observed luminosity between the most and least favorable geometric
orientations (always face-on versus always edge-on hot spots) is only
a factor of $\sim$3 as a consequence of gravitational bending of light
near the neutron star surface. Thus, thermal MSPs should be observable
in X-rays even if they cannot be seen in the radio due to unfavorable
beaming and fall within a relatively narrow range of X-ray
luminosities.  The limiting sensitivity of the available ACIS-S data
is $\sim$$10^{29}$ ergs s$^{-1}$, meaning that typical thermal MSPs
(with luminosities of a few $\times10^{30}$ ergs s$^{-1}$) would be
detected in the combined \textit{Chandra} ACIS-S observations if they
were located in NGC 6397, even in the unlikely event that their two
hot spots are always seen edge-on.  Binary MSPs that eclipse in the
radio tend to have X-ray luminosities $\sim$$10^{31}$ ergs s$^{-1}$
\citep{Stap03,Bog05,Bog06,Els08}. Moreover, the intrabinary shock in
such systems should presumably be visible in X-rays regardless of the
pulsar geometry and the orientation of the binary. Therefore, based on
our current understanding of the X-ray properties of these sources,
\textit{all} MSPs in NGC 6397 should be detected with high
significance in the available data.

%
%
\begin{figure}[t] 
\begin{center}
\includegraphics[width=0.45\textwidth]{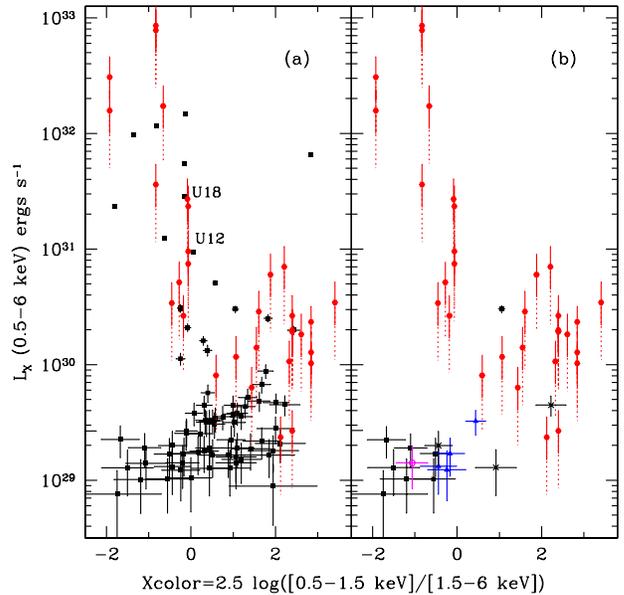}
\caption{(\textit{a}) X-ray luminosity versus color diagram for NGC
  6397. The black points show all 79 sources within the NGC 6397
  half-mass radius (Table 3), for an assumed powerlaw spectrum with
  $\Gamma=2.5$. (\textit{b}) Same as (\textit{a}) but with all sources
  with probable optical identifications removed. The red circles in
  both panels correspond to simulated MSPs based on the current
  Galactic sample of MSPs detected in X-rays (see text for
  details). Sources with no optical counterparts are marked with black
  squares, those with ambiguous or unknown optical counterparts with
  solid blue triangles, the main sequence star candidates with black
  crosses, and the main-sequence turnoff star candidate with the open
  magenta circle. The dotted lines extending below the error bars for
  the MSPs illustrate the effect of the uncertainty in the viewing
  geometry of the X-ray-emmiting polar caps.}
\end{center}
\end{figure}

We have compiled a set of ``template'' MSP X-ray colors and
luminosities using the cluster and field MSPs that have been detected
in X-rays to date. These include the 11 MSPs listed in Table 1 of
\citet{Pavlov07}, the 15 MSPs in the globular clusters 47 Tuc with
unconfused spectra \citep{Bog06}, one MSP in M71 \citep{Els08}, two
MSPs in Ter 5 \citep{Hein06}, and the recently discovered PSR
J1023+0038 in the field of the Galaxy \citep{Arch09}. Using the
measured spectral properties of these MSP we calculate their X-ray
luminosities and colors as they would appear in NGC 6397 using
$N_H=2\times10^{21}$ cm$^{-2}$ and $D=2.4$ kpc and compare them to the
cluster X-ray source population (see Figure 6). We define the color as
${\rm Xcolor} = 2.5\log({\rm counts}[0.3-1.5~{\rm keV}])/{\rm
  counts}[1.5-6~{\rm keV}])$ and consider the absorbed luminosity in
the 0.5--6 keV band. To estimate the uncertainties in the derived MSP
luminosities, we have considered a nominal 30\% error in the pulsar
distances plus the uncertainty due to the range of possible distances
to NGC 6397.  Furthermore, we have taken into account the lack of
information regarding the viewing angles of the X-ray-emitting polar
caps of any thermal MSPs in NGC 6397 (represented by the dotted lines
in Figure 6).  High-quality X-ray spectra of nearby thermal MSPs show
at least two thermal components and a much fainter hard tail above
$\sim$2 keV. For crude photon statistics, the X-ray emission of these
MSPs resembles a powerlaw with photon index $\Gamma\sim2-3$ in the
0.5--6 keV range. Thus, to estimate the X-ray luminosities of the
sources in NGC 6397 we consider $\Gamma=2.5$. To ascertain the effect
of our choice of spectral model we also used a pure
blackbody with $kT=0.2$ keV and varied $N_{\rm H}=(1-3)\times10^{21}$
cm$^{-1}$. We find that the assumptions regarding the spectral shape
do not affect the conclusions of our analysis.

Using the computed set of MSPs, we attempt to identify MSPs in NGC
6397 without the benefit of radio timing searches \citep[see][for a
similar study for 47 Tuc]{Hein05}. Important constraints can be gained
from optical counterpart identifications using \textit{HST} ACS/WFC
observations of the core of NGC 6397 (H.~Cohn et al. in
preparation). Of the 79 X-ray sources detected within the half-mass
radius of NGC 6397, 64 sources have probable associations in the
optical (see Table 3) predominantly with active binaries (ABs) and
cataclysmic variables (CVs). Additionally, U24 is classified as a
qLMXB based on its X-ray properties \citep{Grind01}. The remaining
sources either have ambiguous/unknown or no optical counterparts.

Solitary MSPs should have virtually no optical
counterparts\footnote{The current best limits on optical emission from
  isolated MSPs come from observations of the nearby PSRs J0030+0451
  ($D\approx300$ pc) and J2124--3358 ($D\approx270$ pc), for which no
  counterparts are found down to B, V, and R magnitudes of ~$27-28$
  \citep{Kop03,Mig04}.}.  Therefore, any X-ray source without an
optical counterpart is a viable MSP candidate.  Binary MSPs both in
clusters and the field are commonly bound to low-mass He white dwarfs
or very-low-mass (brown-dwarf-like) companions ($\sim$0.03
M$_{\odot}$) such as in the canonical "black-widow" system PSR
B1957+20 \citep{Fru88}.  For NGC 6397, \citet{Strick09} have found
that none of the 41 He white dwarfs identified optically have X-ray
counterparts, implying that they are not partnered with MSPs. The
optical properties of black widow systems are not well established as
only two objects have been studied in any detail, PSRs B1957+20
\citep{Rey07} and J2051--0827 \citep{Stap01}.
Nonetheless, these systems tend to have $L_X\gtrsim10^{31}$ ergs
s$^{-1}$ \citep{Stap03,Bog06,Els08} so the lack of unidentified
sources with comparable X-ray luminosities implies the absence of
black widow systems in NGC 6397.  Moreover, this indicates that aside
from J1740--5340 and U18, there are no additional peculiar interacting
binaries in this cluster either.  Only a single source without an
optical counterpart (U113 in Table 3) has X-ray colors and luminosity
(${\rm Xcolor}=1.05$ and $\log L_X=30.5$) consistent with those of a
typical MSP (Figure~6).  The sole source with an ambiguous/unknown
optical counterpart that has soft colors, U16, as
well as U41 and U112, which appears to hav probable main-sequence optical
counterparts, are also possible MSP candidates. Based on these
findings, we obtain a limit of $\le$6 MSPs in NGC 6397.

\section{DISCUSSION}

\subsection{PSR J1740--5340}

Optical observations of the PSR J1740--5340 system
\citep{Ferr03,Sabbi03} have revealed the presence of ellipsoidal
variations as well as a stream of gas protruding from the 1.6
$R_{\odot}$ secondary star through the L1 point implying that the
companion is Roche-lobe overflowing. However, the fact that PSR
J1740--5340 is observed as a radio pulsar implies that this gas stream
never reaches the underlying NS and is instead swept back and probably
expelled from the binary system by the relativistic particle wind from
the MSP. The presence of this swept-back gas stream is suggested by
the very unusual H$\alpha$ emission line profile \citep{Sabbi03}. This
gas is likely the cause for the irregular radio eclipses the MSP
exhibits over a wide range of orbital phases.

As seen in ``black-widow'' systems \citep{Stap03,Bog06,Els08} and
``exchanged'' MSP systems \citep{Bog05} the interaction between the
pulsar wind and material from the companion star should result in a
shock front where the ram pressure of the wind balances the pressure
of the infalling gas leading to X-ray emission due to particle
acceleration \citep{Arons93}.  In this scenario, synchrotron emission
is expected to be the primary energy loss mechanism in the resulting
shock wave, given that it occurs in a relatively strong magnetic
field. In the PSR J1740--5340 binary, for a shock situated
approximately at the L1 point and an isotropic pulsar wind, the pulsar
magnetic field at this distance ($\approx$$4\times10^{11}$ cm)
immediately upstream from the shock is $B_1\approx2-3$ G or
$B_1\approx0.1$ G , implying a post-shock field of $B_2=3B_1\sim6-9$ G
or $B_2\sim0.3$ G \citep{Arons93}. The two values correspond to the
two possible cases of a magnetically dominated shock ($\sigma\gg 1$,
where $\sigma$ is the magnetic to kinetic energy flux ratio) or a
kinetic energy dominated shock ($\sigma=0.003$ as in the Crab pulsar).
This implies that the X-ray emission from PSR J1740--5340 source is
most likely non-thermal in origin and not from a thermal plasma.

An intrabinary shock origin of the X-rays from J1740--5340 is also
suggested by the fact that the derived spectral parameters are quite
similar to those of the canonical ``black-widow'' pulsar, PSR B1957+20
\citep{Stap03}, the peculiar MSP-main-sequence binary in 47 Tuc, PSR
J0024--7204W \citep{Edm02,Bog05}, and the recently identified radio
MSP in the field of the Galaxy, PSR J1023+0038 \citep[][and references
therein]{Arch09}, which also appears to be bound to a non-degenerate
(``main-sequence-like'') star. Moreover, the shape of the X-ray
modulations at the binary period are quite similar to that of
J0024--7204W \citep[see Figure 1 in][]{Bog05}.  This variability can
be attributed to an occultation of the shocked material by the
secondary star. This scenario is supported by the compelling (albeit
marginal) evidence for a softening of the spectrum at
$\phi\approx0.25$.  The constraint on the system inclination
\citep[$47^{\circ}-56^{\circ}$][]{Ferr03} suggests that the line of
sight to the pulsar is never obstructed by the companion. Therefore,
when the shock is occulted at $\phi\approx0.25$ (either partially or
totally), the soft thermal radiation from the MSP surface provides a
larger contribution to the total X-ray emission, resulting in a
softening of the observed spectrum. In principle, detailed
orbital-phase-resolved X-ray observations would permit disentanglement
of the thermal polar cap emission and the non-thermal shock emission.

Note that the moderate spin-down luminosity\footnote{As discussed by
  \citet{DAmico01}, the effect of cluster acceleration on the measured
  pulsar spindown rate is neglegible.} of
$\dot{E}\approx3.3\times10^{34}$ ergs s$^{-1}$ \citep{Bass04b} does not
favor a magnetospheric origin of the observed non-thermal X-rays
considering that most other MSPs with comparable values of $\dot{E}$
exhibit much softer, predominantly thermal spectra \citep[see,
e.g.,][]{Bog06,Zavlin06,Bog09}. Furthermore, for pulsed non-thermal
emission, it is difficult to explain the observed variations in the
total flux and the softening of the spectrum at $\phi\approx0.25$.

\subsection{U18: A Hidden MSP?}

The X-ray and optical properties of U18 are consistent with those of a
binary containing a rotation-powered pulsar wind interacting with
material from the secondary star.  It is interesting to note the
significantly larger luminosity compared to J1740--5340 and other
interacting MSP systems. This could indicate either a much larger
shock region, an enhanced density of radiating particles, and/or a
more energetic pulsar wind.  If this binary does indeed harbor a MSP,
it may be difficult to detect at radio frequencies due to the large
quantities of gas present within and around the binary. This gas may
render the pulsar perpetually eclipsed at radio frequencies. Thus, U18
may belong to the class of so-called ``hidden'' MSPs \citep{Tav91}
making a confirmation of its true nature quite difficult. The
detection of pulsed X-ray emission from the pulsar itself may also be
difficult due to the dominant X-ray flux from the intrabinary shock
and the \textit{a priori} unknown pulsar spin period.

We note that, at present, we cannot strictly rule out the possibility
that U18 is a cataclysmic variable, such as an intermediate polar
(IP), instead of an MSP. These objects also generally exhibit hard
X-ray spectra and X-ray luminosities comparable to that of U18. In
addition, the IP AKO9 in 47 Tuc \citep{Hein05} appears to have a ``red
straggler'' or ``sub-subgiant'' companion as well.  In order to unveil
the true nature of the U18 system, further optical observations,
optimized for variability study, are needed. The cadence of our
\textit{HST} ACS/WFC optical dataset (10 single-orbit observations
spaced $\sim$1 month apart, see H.~Cohn et al.~2009, in preparation)
is not suitable for a reliable determination of the binary period of
this system.

\subsection{The Millisecond Pulsar Population of NGC 6397}

The analysis in \S5 suggests that NGC 6397 most likely harbors $1-6$
MSPs.  It is of interest to compare these limits to the expected
number of MSPs in this cluster, scaling by the stellar encounter rate
\citep[$\Gamma$,][]{Verb87} from other clusters with detected
MSPs\footnote{We note that the numbers of (quiescent) LMXBs in
  globular clusters, likely progenitors of MSPs, appear to scale with
  $\Gamma$ \citep{Poo06,Hein06}.}.  \citet{Poo03} integrate collision
rates over King models out to the half-mass radius, finding a
collision rate for NGC 6397 $1/74$ that of 47 Tuc.  Other analyses
\citep[e.g.][]{Hein03,Bass08}, using a simpler formalism
($\Gamma=\rho_0^{1.5} r_c^2$, where $\rho_0$ is the central density
and $r_c$ is the core radius of the cluster), find collision rates for
NGC 6397 roughly $1/10$ that of 47 Tuc.  As 47 Tuc has 23 known MSPs,
and has been suggested to have a total of $\sim$25 MSPs
\citep{Hein05}, the two approaches suggest either $<1$ or $\sim$2 MSPs
should be present in NGC 6397.  Comparisons with the number of MSPs in
other clusters, such as Terzan 5 \citep{Ran05} give similar results.
However, $\Gamma$ comparisons seem to underpredict the number of X-ray
sources in NGC 6397 \citep{Poo03}, indicating differences in the X-ray
binary production processes in core-collapsed clusters such as NGC
6397.  For instance, \citet{Fre08} has suggested that continuing
globular cluster collapse means that the values of $\Gamma$ of
King-model clusters during the epoch of X-ray binary formation were a
factor of 3 smaller.  In any case, the predicted numbers of MSPs in
NGC 6397 from any of the above comparisons are generally in agreement
with our estimate of $1-6$ MSPs in NGC 6397.

\section{CONCLUSION}

We have presented \textit{Chandra X-ray Observatory} ACIS-S
observations of the nearby core-collapsed globular cluster NGC
6397. The depth of the available data has allowed interesting insight
into the MSP population of this cluster.  For instance, it has
provided a more complete multi-wavelength picture of the PSR
J1740--5340 binary. This peculiar system appears to exhibit
predominantly non-thermal emission modulated at the orbital
phase. Given the pulsar energetics, the variable non-thermal X-ray
emission is indicative of interaction between the relativistic pulsar
wind and material from the companion, with evidence for an occultation
of the resulting shock by the secondary star. Ascertaining the
detailed geometry of the intrabinary shock would require much more
sensitive orbital-phase resolved spectroscopic X-ray and optical
observations of this system than presently available.

Based on its X-ray and optical properties, U18 is a strong candidate
for a MSP, though one that may be perpetually eclipsed at radio
frequencies. As noted by \citet{Tav91} such MSPs may be relatively
common both in globular clusters and in the field of Galaxy.
Moreover, these sources may be mis-classified as LMXBs or CVs based on
their X-ray to optical flux ratio alone \citep[see in
particular][]{Homer06,Arch09}.  Further detailed multi-wavelength
observations of PSR J1740--5340, U18, and similar systems may reveal
more information concerning the physics of the shock, which, may, in
principle offer insight into the properties of MSP winds,
collisionless relativistic shocks, and particle acceleration
mechanisms.  Moreover, this and similar MSP binaries provide important
clues about the behavior of accreting neutron stars transitioning to
rotation power, which are believed to also contain an active radio
pulsar and an overflowing companion \citep{Camp04,Hein07,Hein09}.

The deep X-ray observations of NGC 6397 supplemented by optical
observations, also provide insight into the MSP population of this
cluster. Specifically, if the X-ray luminosity function of the NGC
6397 MSP population resembles that of the Galactic sample of MSPs,
then this cluster contains no more than $\sim$6 MSPs. This range of
values is generally consistent with those inferred based on scaling of
stellar encounter rates in globular clusters.

\acknowledgements 
This work was funded in part by NASA
\textit{Chandra} grant GO7-8033A, awarded through the Harvard College
Observatory.  S.~B. is supported in part by a Canadian Institute for
Advanced Research Junior Fellowship.  The research in this paper has
made use of the NASA Astrophysics Data System (ADS).

\LongTables
\footnotesize
\begin{deluxetable*}{rccccrrrc}[]
\tabletypesize{\footnotesize} 
\tablecolumns{10} 
\tablewidth{0pc}
\tablecaption{X-ray Sources in NGC 6397}

\tablehead{ \colhead{} & \colhead{} & \colhead{} & \colhead{} & \colhead{} & \colhead{Soft} & \colhead{Hard} & \colhead{} & \colhead{} \\ \colhead{U} & \colhead{CXOGlb J} & \colhead{$\alpha$} & \colhead{$\delta$} & \colhead{$r_{95\%}$} & \colhead{Counts\tablenotemark{d}} & \colhead{Counts\tablenotemark{d}} & \colhead{Flux\tablenotemark{e}} & \colhead{ID} \\ \colhead{Name\tablenotemark{a}} & \colhead{Name} & \colhead{(J2000.0)\tablenotemark{b}} & \colhead{(J2000.0)\tablenotemark{b}} & \colhead{($\arcsec$)\tablenotemark{c}} & \colhead{(0.5--1.5 keV)} & \colhead{(1.5--6 keV)} & \colhead{(0.5--6 keV)} & \colhead{Type\tablenotemark{f}} \\ \colhead{(1)} & \colhead{(2)} & \colhead{(3)} & \colhead{(4)} & \colhead{(5)} & \colhead{(6)} & \colhead{(7)} & \colhead{(8)} }
\startdata
5	& 174054.5--534044	        &       17      40  54.531 	&	--53	40   44.57  & 	0.33	&	167	&	31	&	41.7	&	AB?	\\
7	& 174052.8--534121      	&	17	40  52.832 	&	--53	41   21.77  & 	0.32	&	252	&	150	&	84.6	&	CV	\\
10	& 174048.9--533948      	&	17	40  48.978 	&	--53	39   48.62  & 	0.29	&	283	&	1504	&	390.8	&	CV	\\
11	& 174045.7--534041      	&	17	40  45.781 	&	--53	40   41.52  & 	0.32	&	72	&	55	&	26.7	&	CV	\\
12	& 174044.6--534041      	&	17	40  44.621 	&	--53	40   41.60  & 	0.30	&	382	&	363	&	157.3	&	MSP	\\
13	& 174044.0--534039      	&	17	40  44.084 	&	--53	40   39.17  & 	0.31	&	107	&	135	&	51.4	&	CV	\\
14	& 174043.3--534155      	&	17	40  43.328 	&	--53	41   55.46  & 	0.40	&	17	&	10	&	5.7	&	AB	\\
15	& 174042.9--534033      	&	17	40  42.910 	&	--53	40   33.81  & 	0.34	&	42	&	9	&	11.2	&	AB	\\
16	& 174042.6--534215      	&	17	40  42.636 	&	--53	42   15.24  & 	0.42	&	15	&	10	&	5.4	&	?	\\
17	& 174042.6--534019      	&	17	40  42.651 	&	--53	40   19.30  & 	0.28	&	5358	&	6114	&	2458.0	&	CV	\\
18	& 174042.6--534027      	&	17	40  42.606 	&	--53	40   27.62  & 	0.29	&	995	&	1142	&	470.2	&	MSP?	\\
19	& 174042.3--534028      	&	17	40  42.306 	&	--53	40   28.70  & 	0.28	&	2777	&	5934	&	1927.3	&	CV	\\
21	& 174041.8--534021      	&	17	40  41.830 	&	--53	40   21.37  & 	0.29	&	1974	&	2279	&	901.4	&	CV	\\
22	& 174041.7--534029      	&	17	40  41.701 	&	--53	40   29.00  & 	0.31	&	104	&	131	&	50.3	&	CV	\\
23	& 174041.5--534019      	&	17	40  41.597 	&	--53	40   19.30  & 	0.28	&	1677	&	5915	&	1609.5	&	CV	\\
24	& 174041.4--534004      	&	17	40  41.468 	&	--53	40   04.43  &   0.28	&	4797	&	353	&	1088.8	&	qLMXB	\\
25	& 174041.2--534025      	&	17	40  41.237 	&	--53	40   25.79  & 	0.33	&	39	&	49	&	18.7	&	CV?	\\
28	& 174038.9--533951      	&	17	40  38.904 	&	--53	39   51.09  & 	0.30	&	351	&	629	&	206.1	&	GLX	\\
31	& 174034.2--534115      	&	17	40  34.202 	&	--53	41   15.28  & 	0.37	&	26	&	18	&	9.5	&	CV	\\
41	& 174045.0--533955      	&	17	40  45.008 	&	--53	39   55.21  & 	0.36	&	31	&	4	&	7.5	&	MS?	\\
42	& 174043.0--533831      	&	17	40  43.059 	&	--53	38   31.29  & 	0.34	&	141	&	15	&	33.0	&	AB	\\
43	& 174040.5--534022      	&	17	40  40.543 	&	--53	40   22.79  & 	0.36	&	25	&	10	&	7.4	&	AB	\\
60	& 174047.8--534128      	&	17	40  47.807 	&	--53	41   28.40  & 	0.41	&	10	&	11	&	4.4	&	CV	\\
61	& 174045.2--534028      	&	17	40  45.223 	&	--53	40   28.60  & 	0.32	&	79	&	85	&	34.8	&	CV?	\\
62	& 174030.4--533917      	&	17	40  30.422 	&	--53	39   17.47  & 	0.48	&	12	&	5	&	3.7	&	AB?	\\
63	& 174031.6--533846      	&	17	40  31.663 	&	--53	38   46.36  & 	0.42	&	31	&	7	&	8.0	&	AB?	\\
65	& 174037.5--533917      	&	17	40  37.558 	&	--53	39   17.85  & 	0.43	&	14	&	3	&	3.6	&	AB?	\\
66	& 174038.9--533849      	&	17	40  38.918 	&	--53	38   49.80  & 	0.48	&	9	&	4	&	2.7	&	AB?	\\
67	& 174040.0--534016      	&	17	40  40.065 	&	--53	40   16.59  & 	0.41	&	8	&	6	&	3.0	&	AB	\\
68	& 174040.7--533832      	&	17	40  40.730 	&	--53	38   32.63  & 	0.44	&	1	&	20	&	4.4	&	?	\\
69	& 174040.8--534017      	&	17	40  40.867 	&	--53	40   17.17  & 	0.37	&	15	&	9	&	5.1	&	AB	\\
70	& 174041.6--534033      	&	17	40  41.693 	&	--53	40   33.33  & 	0.32	&	59	&	41	&	22.1	&	AB	\\
73	& 174042.6--533928      	&	17	40  42.681 	&	--53	39   28.73  & 	0.38	&	22	&	8	&	6.3	&	AB	\\
75	& 174043.6--534030      	&	17	40  43.666 	&	--53	40   30.61  & 	0.39	&	9	&	10	&	4.2	&	AB	\\
76	& 174043.8--534116      	&	17	40  43.818 	&	--53	41   16.32  & 	0.38	&	21	&	7	&	5.9	&	AB	\\
77	& 174044.1--534211      	&	17	40  44.125 	&	--53	42   11.48  & 	0.42	&	18	&	7	&	5.2	&	AB?	\\
79	& 174046.4--534004      	&	17	40  46.409 	&	--53	40   04.05  &   0.38	&	18	&	9	&	5.9	&	AB	\\
80	& 174046.4--534156      	&	17	40  46.455 	&	--53	41   56.50  & 	0.39	&	20	&	15	&	7.4	&	CV?	\\
81	& 174046.4--534115      	&	17	40  46.481 	&	--53	41   15.44  & 	0.38	&	15	&	11	&	5.5	&	AB	\\
82	& 174048.5--533939      	&	17	40  48.537 	&	--53	39   39.53  & 	0.37	&	31	&	9	&	8.6	&	AB	\\
83	& 174049.6--534043      	&	17	40  49.615 	&	--53	40   43.02  & 	0.46	&	5	&	6	&	2.4	&	CV?	\\
84	& 174054.8--534019      	&	17	40  54.807 	&	--53	40   19.79  & 	0.40	&	32	&	5	&	7.8	&	AB?	\\
86	& 174037.4--534147      	&	17	40  37.473 	&	--53	41   47.24  & 	0.35	&	56	&	11	&	14.6	&	AB?	\\
87	& 174042.8--534026      	&	17	40  42.877 	&	--53	40   26.45  & 	0.37	&	14	&	10	&	5.3	&	AB	\\
88	& 174042.8--534023      	&	17	40  42.863 	&	--53	40   23.43  & 	0.37	&	20	&	8	&	6.2	&	AB	\\
89	& 174043.6--534024      	&	17	40  43.613 	&	--53	40   24.60  & 	0.41	&	11	&	3	&	3.1	&	AB	\\
90	& 174041.7--534014      	&	17	40  41.779 	&	--53	40   14.42  & 	0.36	&	26	&	8	&	7.2	&	AB	\\
91	& 174042.4--534041      	&	17	40  42.430 	&	--53	40   41.65  & 	0.40	&	6	&	9	&	3.3	&	MS	\\
92	& 174043.9--534035      	&	17	40  43.916 	&	--53	40   35.39  & 	0.40	&	11	&	4	&	3.2	&	AB?	\\
93	& 174042.3--534046      	&	17	40  42.393 	&	--53	40   46.62  & 	0.42	&	6	&	7	&	2.8	&	?	\\
94	& 174042.8--534049      	&	17	40  42.868 	&	--53	40   49.07  & 	0.40	&	14	&	2	&	3.4	&	AB	\\
95	& 174040.3--534044      	&	17	40  40.320 	&	--53	40   44.58  & 	0.47	&	4	&	4	&	1.7	&	AB?	\\
96	& 174039.0--534023      	&	17	40  39.097 	&	--53	40   23.09  & 	0.42	&	11	&	2	&	2.7	&	AB	\\
97	& 174043.9--534005      	&	17	40  43.918 	&	--53	40   05.90  &   0.44	&	4	&	6	&	2.2	&	?	\\
98	& 174040.9--534058      	&	17	40  40.994 	&	--53	40   58.40  & 	0.41	&	12	&	2	&	3.0	&	AB	\\
99	& 174046.4--534030      	&	17	40  46.431 	&	--53	40   30.40  & 	0.50	&	6	&	1	&	1.5	&	AB	\\
100	& 174038.2--534046      	&	17	40  38.201 	&	--53	40   46.55  & 	0.45	&	6	&	4	&	2.1	&	AB	\\
101	& 174045.3--534101      	&	17	40  45.399 	&	--53	41   01.40  &   0.39	&	19	&	3	&	4.6	&	AB?	\\
102	& 174038.8--533943      	&	17	40  38.845 	&	--53	39   43.12  & 	0.40	&	11	&	9	&	4.2	&	AB	\\
103	& 174035.6--534012      	&	17	40  35.698 	&	--53	40   12.56  & 	0.44	&	9	&	3	&	2.5	&	AB	\\
104	& 174043.1--533929      	&	17	40  43.124 	&	--53	39   29.04  & 	0.44	&	5	&	8	&	2.8	&	none	\\
105	& 174036.5--534107      	&	17	40  36.521 	&	--53	41   07.85  &   0.44	&	8	&	5	&	2.7	&	AB	\\
106	& 174043.7--533917      	&	17	40  43.737 	&	--53	39   17.52  & 	0.51	&	3	&	5	&	1.7	&	none	\\
107	& 174034.1--534017      	&	17	40  34.115 	&	--53	40   17.01  & 	0.39	&	16	&	10	&	5.4	&	AB	\\
108	& 174052.0--533948      	&	17	40  52.099 	&	--53	39   48.25  & 	0.46	&	9	&	6	&	3.2	&	GLX?	\\
109	& 174052.7--534052      	&	17	40  52.728 	&	--53	40   52.88  & 	0.50	&	8	&	3	&	2.3	&	AB?	\\
110	& 174033.4--533916      	&	17	40  33.455 	&	--53	39   16.83  & 	0.50	&	8	&	3	&	2.6	&	AB	\\
111	& 174029.8--534026      	&	17	40  29.845 	&	--53	40   26.99  & 	0.45	&	3	&	14	&	3.7	&	none	\\
112	& 174050.3--533906      	&	17	40  50.374 	&	--53	39   06.00  &   0.53	&	7	&	3	&	2.1	&	MS	\\
113	& 174042.7--534020      	&	17	40  42.764 	&	--53	40   20.76  & 	0.31	&	161	&	61	&	50.0	&	none	\\
114	& 174043.4--534034      	&	17	40  43.469 	&	--53	40   34.34  & 	0.45	&	4	&	5	&	2.0	&	?	\\
116	& 174042.2--534019      	&	17	40  42.236 	&	--53	40   19.97  & 	0.36	&	15	&	14	&	6.3	&	AB	\\
117	& 174042.1--534025      	&	17	40  42.153 	&	--53	40   25.56  & 	0.41	&	7	&	6	&	2.9	&	AB	\\
118	& 174041.5--534015      	&	17	40  41.576 	&	--53	40   15.88  & 	0.43	&	3	&	8	&	2.3	&	MSTO	\\
119	& 174041.2--534019      	&	17	40  41.261 	&	--53	40   19.35  & 	0.40	&	4	&	11	&	3.2	&	none	\\
120	& 174046.5--534015      	&	17	40  46.517 	&	--53	40   15.64  & 	0.53	&	6	&	0	&	1.3	&	AB?	\\
121	& 174033.6--533934      	&	17	40  33.631 	&	--53	39   34.96  & 	0.50	&	2	&	8	&	2.1	&	none	\\
122	& 174047.9--533924      	&	17	40  47.903 	&	--53	39   24.83  & 	0.58	&	1	&	5	&	1.3	&	none	\\
123	& 174049.6--533845      	&	17	40  49.621 	&	--53	38   45.93  & 	0.60	&	2	&	6	&	1.7	&	none  \\	
\enddata
\tablenotetext{a}{Source names based on the X-ray source list from
  \citet{Grind01}.}  
\tablenotetext{b}{J2000.0 {\tt acis\_extract}
  derived positions of each source boresighted to the frame of the
  \textit{HST} ACS dataset.} 
\tablenotetext{c}{The 95\%
  positional uncertainty radius based on \citet{Hong05}, in units of arcseconds.}
\tablenotetext{d}{Background-subtracted source counts in the 0.3--1.5
  and 1.5--6 keV bands.}  
\tablenotetext{e}{Source flux in the 0.5--6
  keV band in units of $10^{-16}$ ergs cm$^{-2}$ s$^{-1}$ assuming a
  power-law model with spectral photon index $\Gamma=2.5$.}
\tablenotetext{f}{Probable source identifications based on optical
  counterpart properties (from H.~Cohn et al., in preparation). AB is
  an active binary, CV a cataclysmic variable, GLX a background
  galaxy, qLMXB a quiescent low mass X-ray binary,
  MS a main sequence star, MSTO a main sequence turn-off star, and
  MSP a millisecond pulsar. Sources marked with an additional ``?''
  are probable but less certain classifications, those with only ``?''
  have ambiguous or unknown, and those labeled ``none'' have no
  optical counterparts.}
\end{deluxetable*}

\end{document}